\documentclass[journal]{IEEEtran}
\IEEEoverridecommandlockouts

\usepackage{multirow}
\usepackage{cite}
\usepackage{amsmath,amssymb,amsfonts}
\usepackage{commath}
\usepackage{algorithmic}
\usepackage{graphicx}
\usepackage{textcomp}
\usepackage{xcolor}
\usepackage{array}
\usepackage{stfloats}
\def\BibTeX{{\rm B\kern-.05em{\sc i\kern-.025em b}\kern-.08em
    T\kern-.1667em\lower.7ex\hbox{E}\kern-.125emX}}
\begin{document}

\title{Investigating A Geometrical Solution to the Vergence-Accommodation Conflict for Targeted Movements in Virtual Reality
\thanks{This work was supported through grants from the Social Sciences and Humanities Research Council of Canada (SSHRC), the Canada Research Chair Program, the Natural Sciences and Engineering Research Council of Canada (NSERC), the Canada Foundation for Innovation (CFI), and the Ontario Ministry for Research and Innovation. The authors wish to thank Madeleine Corredoura and Colin Dolynski for assisting with the data collection.}
}

\author{
Xiaoye Michael Wang, Matthew Prenevost, Aneesh Tarun, Ian Robinson
\\
Michael Nitsche, Gabby Resch, Ali Mazalek, Timothy N. Welsh
\thanks{This work involved human subjects in its research. Approval of all ethical and experimental procedures and protocols was granted by the University of Toronto Research Ethics Board under Protocol \#00042432 and performed in line with the Regulations and Policies of the University of Toronto.}
\thanks{Xiaoye Michael Wang, Matthew Prenevost, and Timothy N. Welsh are with the University of Toronto, Toronto, Canada (email: michaelwxy.wang@utoronto.ca, matthew.prenevost@mail.utoronto.ca, t.welsh@utoronto.ca)}
\thanks{Aneesh Tarun, Ian Robinson, and Ali Mazalek are with Toronto Metropolitan University, Toronto, Canada (email: aneesh@torontomu.ca, robinson.ian@torontomu.ca, mazalek@torontomu.ca)}
\thanks{Michael Nitsche is with Georgia Tech University, Atlanta, USA (email: mn81@gatech.edu)}
\thanks{Gabby Resch is with Ontario Tech University, Oshawa, Canada (email: Gabby.Resch@ontariotechu.ca)}
}

\maketitle

\begin{abstract}

While virtual reality (VR) holds significant potential to revolutionize digital user interaction, how visual information is presented through VR head-mounted displays (HMDs) differs from naturalistic viewing and interactions in physical environments, leading to performance decrements. One critical challenge in VR development is the vergence-accommodation conflict (VAC), which arises due to the intrinsic constraints of approximating the natural viewing geometry through digital displays. Although various hardware and software solutions have been proposed to address VAC, no commercially viable option has been universally adopted by manufacturers. This paper presents and evaluates a software solution grounded in a vision-based geometrical model of VAC that mediates VAC’s impact on movement in VR. This model predicts the impact of VAC as a constant offset to the vergence angle, distorting the binocular viewing geometry that results in movement undershooting. In Experiment 1, a 3D pointing task validated the model’s predictions and demonstrated that VAC primarily affects online movements involving real-time visual feedback. Experiment 2 implemented a shader program to rectify the effect of VAC, improving movement accuracy by approximately 30\%. Overall, this work presented a practical approach to reduce the impact of VAC on HMD-based manual interactions, enhancing the user experience in virtual environments.
\end{abstract}


\section{Introduction}


Modern VR devices use stereoscopic displays and “near” real-time marker-based or vision-based motion tracking to simulate the visual and interaction experience of the physical environment \cite{wang2023relating, wang2024integration}. Despite having the term "reality" in its name, visual perception and visually guided actions in VR have been shown to be different from those in the physical environment. These differences include distorted depth perception \cite{creem2023perceiving, wang_geometry_2024, renner_perception_2013, harris_virtually_2019}, altered sensorimotor control \cite{wang_prolonged_2024, wang_mixed_nodate, gagnon_effect_2021, gagnon_estimating_2021}, and henceforth. Differences in behavior in VR could be attributed to many perturbations that the VR HMD imposes on the users, such as increased weight \cite{yan2019effects}, restricted field of view \cite{teixeira2021effects}, motion-to-photon delay \cite{warburton_measuring_2022}, and a lack of haptic feedback \cite{gibbs2022comparison} (see \cite{kelly_distance_2022, creem2023perceiving} for a review).\\
\indent
One persistent technical challenge that VR faces is the vergence-accommodation conflict (VAC) \cite{batmaz_effect_2022, batmaz_re-investigating_2023, bingham_accommodation_2001, hoffman_vergenceaccommodation_2008, hussain2023improving, mon-williams_recent_1999, spiegel2024vergence, wann_natural_1995, wang_geometry_2024}. When looking at an object with both eyes, the eyes rotate in opposite directions toward the object, a process termed \textit{vergence}. Vergence ensures the images of the object would fall on corresponding locations on each eye’s retina. The angle between the two eyes, known as the vergence angle, corresponds to the object’s distance through triangulation based on oculomotor signals (i.e., the eyes’ angles; however, see also \cite{linton2023minimal}). At the same time, the lens of each eye changes shape in a process termed \textit{accommodation}. Accommodation ensures the light projected to the retina is sharp in focus. Like vergence, the accommodative state of the eyes also correlates with the object’s distance from the observer. In a natural viewing condition in a physical environment, vergence and accommodation are congruent, both specifying the same depth (Fig. \ref{fig:vac_demo}a). However, when wearing a VR HMD, the eyes accommodate to a fixed distance due to the constraints imposed by the screens, while still converge freely to fixate on different objects in the environment (Fig. \ref{fig:vac_demo}b). This discrepancy between vergence (constantly changing) and accommodation (relatively static) disrupts the coupling between the two processes, which has been shown to impact a variety of aspects of user experience in VR, such as user discomfort and fatigue \cite{hoffman_vergenceaccommodation_2008, lambooij_visual_2009}, motion sickness and cybersickness \cite{porcino2020minimizing}, time-to-focus \cite{spiegel2024vergence}, distance perception \cite{bingham_accommodation_2001, kramida_resolving_2015, singh_effect_2018, swan_matching_2015}, and motor planning and control \cite{batmaz_effect_2022, batmaz_re-investigating_2023}. In use scenarios such as surgical training, the impact of VAC becomes particularly critical, in which a few centimeters of error could lead to serious consequences. This poses a substantial challenge for VR-based training in fields where precision is vital, highlighting the need for continued research and development in mitigating the effects of VAC.

\subsection{Contributions}

The current work uses a behavioral approach to understand VAC and alleviate its impact on motor performance. A previous study proposed a geometrical model that describes the impact of VAC on distance perception, suggesting that VAC draws the user's vergence inward due to the fixed accommodative distance \cite{wang_geometry_2024}. In turn, the increased vergence angle perturbs the binocular viewing geometry, which results in distance underestimation and movement undershooting in a manual pointing task. The first user study reported in the current work validated the geometrical model using a manual pointing task and explored the perceptuomotor mechanisms underlying movement inaccuracy. Based on this model, a shader-based software solution was developed to improve the accuracy of targeted movements. Finally, a second user study evaluated the effectiveness of this solution. Overall, the present work offers the following key contributions to the field:

\begin{itemize}
    \item \textbf{A more nuanced understanding of the impact of VAC on targeted movements:} This paper addresses the specific contributions of binocular vision on guiding 3D pointing. The first user study reported in this paper used both online guidance (with real-time visual feedback) and feedforward (memory-based, without real-time visual feedback) movements. This manipulation showed that VAC primarily affects movements performed with online visual feedback through disparity matching, instead of the target’s perceived distance necessitated in the feedforward condition. This finding suggests that mitigating strategies should prioritize reducing VAC’s effects on real-time binocular disparity processing to improve movement accuracy and precision in VR tasks.
    \item \textbf{A custom shader program that improves targeted movement accuracy despite VAC:} The present work leverages the intrinsic geometry of VAC to develop a transformation algorithm designed to improve the accuracy of targeted movement despite VAC. Implemented as a shader program, this algorithm systematically transforms the rendered virtual environment to offset the effect of movement undershooting due to VAC. Unlike other methods of resolving the VAC issue, this algorithm does not require eye tracking, imposes minimal computational overhead, and can be easily applied to any commercially available VR HMDs.
    \item \textbf{A preliminary validation on the effectiveness of the custom shader:} The second user study demonstrates that the transformation algorithm could offset the distance undershooting due to VAC and improve movement endpoint accuracy by approximately 30\%. Additionally, this study revealed significant individual differences in response to the altered virtual environment and the mediating role of target distance in VAC’s impact on movement accuracy. These insights not only deepen the understanding of the impact of VAC on behavior but also highlight critical areas for future research and potential personalized approaches to optimizing VR performance.
\end{itemize}

\section{Related Work}

This section reviews relevant work regarding the mechanisms of VAC and how it impacts targeted movement in VR. The reviews will discuss a fundamental understanding of binocular vision, followed by the ways through which binocular vision is used to guide online movement. Then, specific issues related to VAC and HMD-based VR are discussed.

\subsection{Binocular Vision}

\begin{figure}
    \centering
    \includegraphics[width=1\linewidth]{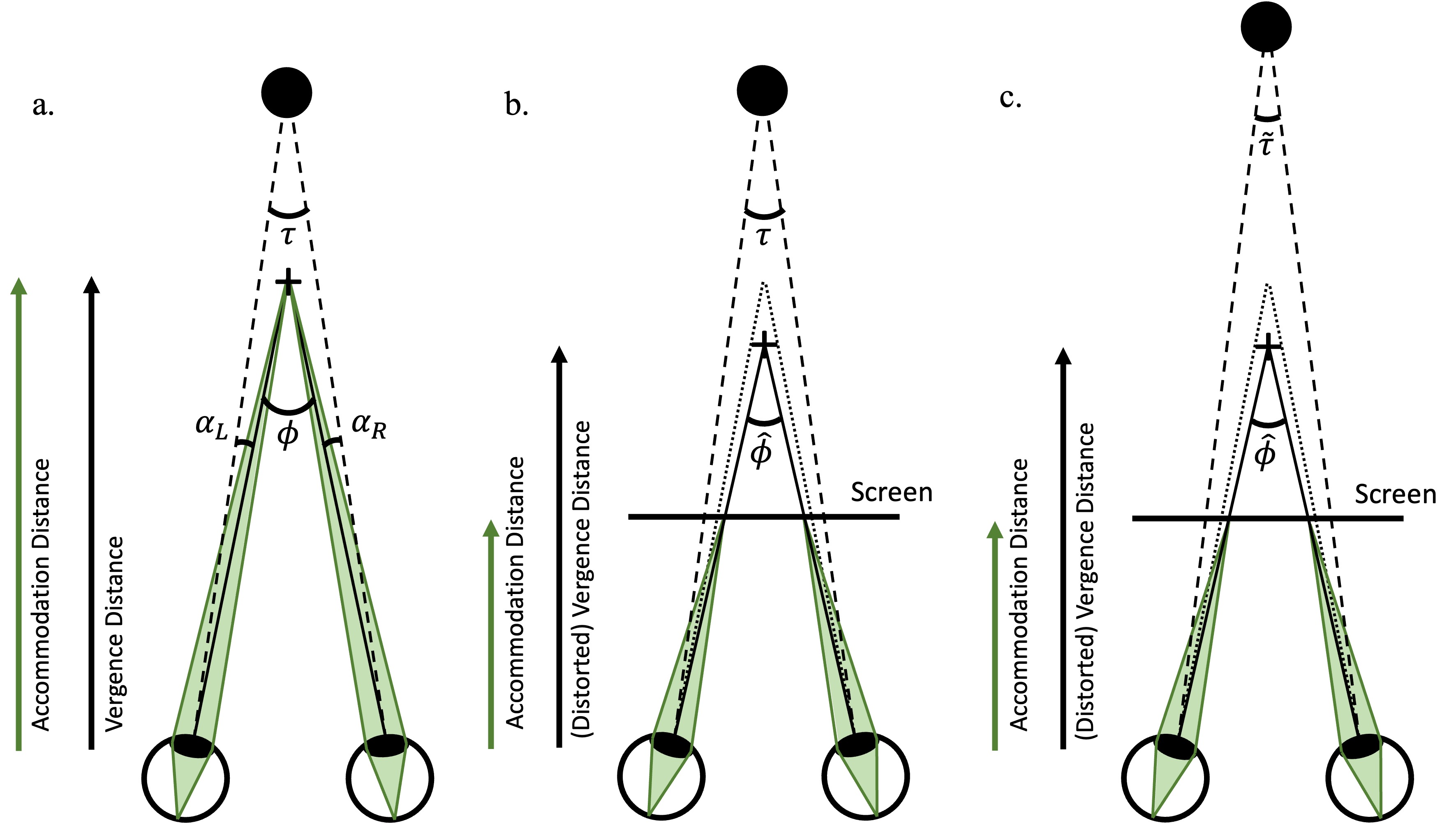}
    \caption{Visualizations of the binocular viewing geometry. (a) The geometry under a normal viewing condition. The binocular disparity associated with the target (black circle) can be derived as the difference of the visual angle of the target relative to the fixation (black cross), $\delta = \alpha_L - \alpha_R$, which is equivalent to the difference between the vergence angle $\phi$ and the visual angle subtended by the target $\tau$, $\delta = \phi - \tau$. Importantly, under a normal viewing condition, vergence (black, solid lines) and accommodation (green areas) are congruent and correspond to the same depth. (b) The perturbed binocular viewing geometry with the mediation of a screen. The screen draws the accommodation to a fixed location, closer than the fixation. As a result, the vergence is also drawn inward, resulting in a larger vergence angle, $\hat{\phi}$. (c) The viewing geometry with the transformation that renders the target further than it is to offset the depth compression induced by the distorted vergence angle.}
    \label{fig:vac_demo}
\end{figure}

Humans have two horizontally separated eyes that, due to their different vantage points, produce slightly different retinal images. The difference between these retinal images is known as binocular disparity, and it provides the impression of depth \cite{julesz_binocular_1964, howard_binocular_1995, lappin2014binocular, qian1997binocular}. Fig. \ref{fig:vac_demo}a illustrates the geometry of binocular disparity (see also Chapter 2 of \cite{howard_binocular_1995}). A target's (black circle) visual angle relative to the fovea in the left eye is $\alpha_L$, negative if the image is to the right of the fixation, and vice versa for that in the right eye, $\alpha_R$, signed in the same way. The binocular disparity is the difference between the two visual angles:
\begin{equation}\label{eq: disparity}
    \delta = \alpha_L - \alpha_R
\end{equation}
Geometrically, this is equivalent to 
\begin{equation}\label{eq: delta}
    \delta = \phi - \tau
\end{equation}
where $\phi$ denotes the visual angle subtended by the fixation between the two eyes (i.e., the vergence angle) and $\tau$ is that by the target. 

To derive depth information from static disparity, additional extra-retinal information, such as the vergence angle and the distance between the eyes (interpupillary distance, or IPD), is required \cite{anderson_solution_2010, backus1999horizontal, banks2001perceiving}. Given Equation \ref{eq: delta}, the target's visual angle is
\begin{equation}
    \tau = \phi - \delta
\end{equation}
Given IPD, the target distance $d$ from the cyclopean eye can be derived as:
\begin{equation}\label{eq: d_from_tau}
    d = \dfrac{IPD / 2}{\tan(\tau / 2)} = \dfrac{IPD / 2}{\tan((\phi - \delta) / 2)}
\end{equation}
Since this extra-retinal information, particularly IPD, is typically unknown to the observer, binocular disparity alone only provides relative distance information. Accurate mapping between disparity and perceived depth requires calibration through targeted movements with visual feedback \cite{bingham1998necessity, bingham_accommodation_2001, bingham2007natural}.

Finally, there is also additional dynamic binocular information, including change of disparity over time (CDOT) and interocular velocity difference (IOVD; \cite{cumming_binocular_1994, shioiri_motion_2000, harris_binocular_2008, nefs_two_2010}). The difference between CDOT and IOVD lies in the order of operations. For CDOT, the static disparity was derived first, followed by taking the disparity's temporal derivative 
\begin{equation}
    \text{CDOT} = \frac{d(\alpha_L - \alpha_R)}{dt}
\end{equation}
For IOVD, the temporal derivative of each eye (i.e., velocity) is derived first, followed by computing the velocity difference between the two eyes:
\begin{equation}
    \text{IOVD} = \frac{d\alpha_L}{dt} - \frac{d\alpha_R}{dt}    
\end{equation}
While mathematically equivalent, CDOT and IVOD rely on distinct neural processing channels \cite{himmelberg_decoding_2020}, resulting in different speed sensitivities \cite{joo_separate_2016, shioiri_motion_2000} and spatial resolutions \cite{czuba_speed_2010}. Functionally, observers with stereo deficiencies may not be able to derive static binocular disparity (e.g., strabismus and amblyopia patients), but could, in turn, rely on the disparity of monocular motion to gain a sense of depth \cite{cardenas-delgado_could_2017}. 

\subsection{The Vergence-Accommodation Conflict}

As discussed in the Introduction, VAC emerges due to the screen-mediated nature of VR technologies. Efforts have been put into resolving VAC through different hardware- and software-based solutions (see \cite{kramida_resolving_2015, hussain2023improving} for a review). One obvious method for resolving VAC is to align the user’s fixation with the HMD’s focal plane by presenting the target at the distance to which the user is accommodating \cite{batmaz_measuring_2023, batmaz_re-investigating_2023, mcanally2024visually}. This technique is commonly employed in vision science. For example, in random dot stereograms that depict 3D environments (e.g., \cite{wang2018large, wang2020symmetry}), points at the same depth plane as the screen exhibit zero disparity, ensuring consistency between vergence and accommodation. While effective, limiting the target’s position to where the user naturally accommodates on a specific device can restrict the immersive and flexible nature of VR experiences. 

On the hardware side, one approach involves adaptive optics, where the focal distance of the display dynamically changes in response to the user’s gaze, allowing the eyes to accommodate at a natural distance \cite{laffont2018verifocal}. Another solution utilizes light field displays, which recreate the way light rays would naturally converge, enabling the eyes to focus correctly at varying distances within the virtual environment \cite{zabels2019ar}. Additionally, multifocal displays present multiple depth planes, offering different focal distances that more closely match vergence \cite{batmaz_effect_2022}. Finally, for the software-oriented solutions, techniques were developed to simulate the depth of field and retinal blur on the screens based on the user’s gaze directions and the scene’s depth structure \cite{hussain2020modelling, hussain2023improving, march2022impact, duchowski2014reducing}. While these technologies hold promise, each solution comes with its own set of challenges, such as increased computational complexity (e.g., \cite{hussain2023improving}), cost (e.g., \cite{laffont2018verifocal}), and the need for further miniaturization to be viable in consumer-grade VR systems (e.g., \cite{batmaz_effect_2022}). Despite advancements, achieving a perfect resolution to VAC remains an ongoing challenge, and a critical area of research in the pursuit of more immersive and comfortable VR experiences.

A common theme among these attempts to resolve VAC is the focus on overcoming the intrinsic hardware limitations of HMD-based VR, aiming to make the user experience in VR as similar as possible to that in the physical environment. However, given the inherent constraints of VR HMDs, it is unlikely that the perceptuomotor experience in VR will perfectly match that of the physical world. Moreover, experience in VR can lead to sensorimotor adaptations to the unique perceptuomotor mappings of the virtual environment \cite{wright_sensorimotor_2014, ebrahimi_effects_2014, ebrahimi_carryover_2015, kohm_objects_2022}. These adaptations may then transfer to the physical environment, potentially impairing performance in real-world tasks \cite{wang_prolonged_2024}. Therefore, it is necessary to identify strategies that could mitigate the effects of VAC. An alternative (and pragmatic) approach is to accept the inevitability of VAC and, through a mechanistic understanding of its behavioral implications, develop ways through which the impact of VAC on movements in the physical and virtual environments can be reduced. This approach prioritizes minimizing the negative impacts on subsequent behaviors in either environment, rather than striving for a seamless equivalence between them \cite{chalmers2003seamful}.

\subsection{The Geometry of the Vergence-Accommodation Conflict} \label{vac_description}

The perceptual impact of VAC can be described using the binocular viewing geometry. Because of VAC, constant accommodation at a closer distance drives vergence inward, resulting in a constant offset, $\beta_\textrm{offset}$, to the vergence angle \cite{bingham_accommodation_2001, swan_matching_2015, singh_effect_2018}. This yields the effective vergence angle of the user as (Fig. \ref{fig:vac_demo}b):
\begin{equation}\label{eq: phihat}
    \hat{\phi} = \phi + \beta_\textrm{offset}
\end{equation}

Iskander and colleagues \cite{iskander2019using} combined eye-tracking in an HMD with a biomechanical model of the eyes to show that the vergence angle mediated by the HMD was greater than that in a natural viewing condition, which, for the HMD used in their experiment (HTC VIVE), was equivalent to a 0.5° to 0.75° offset to the vergence angle. On a behavioral level, this offset could affect where the users perceive the target. Specifically, given the disparity, $\delta$, and perturbed vergence angle, $\hat{\phi}$, the target's effective visual angle could be interpreted as
\begin{equation}\label{eq: tauhat}
    \hat{\tau} = \hat{\phi} - \delta = (\phi + \beta_\textrm{offset}) - \delta
\end{equation} 

Following Equation \ref{eq: d_from_tau}, the perturbed distance based on the target's effective visual angle can be expressed as 
\begin{equation}\label{eq: dhat}
    \hat{d} = \dfrac{IPD / 2}{\tan(\hat{\tau} / 2)}
\end{equation}

Combining Equations \ref{eq: phihat}, \ref{eq: tauhat}, and \ref{eq: dhat} yields:
\begin{equation}
    \hat{d} = \dfrac{IPD / 2}{\tan((\phi + \beta_\textrm{offset}) - \delta )}
\end{equation}

Then, the magnitude of the perceived depth distortion can be computed as the difference between the perceived and actual distance:
\begin{equation}\label{eq: const_err}
    \epsilon = \dfrac{IPD / 2}{\tan((\phi + \beta_\textrm{offset}) - \delta )} - \dfrac{IPD / 2}{\tan(\tau / 2)}
\end{equation}

Wang and colleagues \cite{wang_geometry_2024} fitted this geometrical model to the results of a 3D pointing study in VR. The model described the behavioral errors well, capturing up to 66\% of the variance in the test data. The fitted parameters revealed a constant vergence offset of 0.22° (HTC VIVE Pro) or, equivalently, a fixation distance that was 2.93 cm closer than that specified by the display.

While this vergence offset clearly impacts the binocular viewing geometry, the specific mechanisms by which it translates into errors in 3D pointing remain unclear. Visually guided movements involve both feedforward (i.e., pre-programmed) and online feedback control \cite{woodworth_accuracy_1899, jeannerod1988neural, grierson_goal-directed_2009, elliott_multiple_2017}. Initially, disparity information specifies the relative distance, aiding in the planning and initiation of the feedforward/planned component of the movement. As the hand begins to move and enters the visual field, online control takes over, using the evolving disparity between the hand and the target to minimize the difference between them \cite{anderson_solution_2010, bingham2023stable}. What remains to be determined is whether the VAC-induced errors observed by \cite{wang_geometry_2024} are primarily due to disrupted distance perception during movement planning or the inaccuracies during the online guidance phase.

\section{Experiment 1: Replication}

Experiment 1 was designed to replicate the results from \cite{wang_geometry_2024} with a different setup while introducing two visual feedback conditions: online guidance and feedforward. These conditions manipulate how binocular disparity is used to guide targeted movements in VR. In the online guidance condition, participants could see both the target and their hand throughout the movement, allowing dynamic online movement corrections by continuously matching the disparity between the target and the hand. In the feedforward condition, participants were initially presented with the stimulus but were required to wait to move until after the target disappeared. Thus, in the feedforward condition, the actor relied solely on their visual memory of the target’s position. In this process, participants had to use the initial static disparity to estimate the target’s distance and develop a motor plan accordingly. If the vergence offset impacted distance perception, movement errors would increase in the memory-based feedforward condition. However, if the vergence offset primarily affected dynamic disparity matching, then movement errors would be more pronounced in the online guidance condition, with minimal impact in the feedforward condition.

\subsection{Methods}
\subsubsection{Participants}

Twenty (20) adults participated in Experiment 1 (14 females and 6 males, between 18 and 22 years of age). All participants were right-handed, had normal or corrected-to-normal vision, and were free from any known neurological impairments. Participants provided full and informed consent before taking part in the study either as volunteers or with monetary compensation. 

\subsubsection{Materials and Apparatus}

The experiment was conducted in an immersive virtual environment constructed in Unity. Participants sat at a table and wore an HTC VIVE Pro Eye head-mounted display (HMD) with a resolution of 1440 x 1600 pixels per eye, a combined 110° field of view, and a refresh rate of 90 Hz. The experiment was executed using bmlTUX, a toolkit for designing and running experiments in Unity \cite{bebko_bmltux_2020}.

Manual pointing movements were captured using an opto-electric motion tracking system (Optotrak, Northern Digital Inc., Waterloo, Ontario, Canada). The system recorded the 3D coordinates of an infrared-emitting diode (IRED) attached to the tip of the participant’s right index finger, at a sampling frequency of 250 Hz. A custom Python interface was developed to stream real-time movement data from Optotrak to Unity via the Optotrak Application Programmer’s Interface (OAPI) and a User Datagram Protocol (UDP) socket. The movement data were used to control a virtual hand formed in a pointing posture in the virtual environment with an average latency of 2.74 ms (SD = 2.03 ms).

Participants sat in front of a table in the physical environment while wearing the HMD. Inside the virtual environment (see Fig. \ref{fig:vr_transformation_demo}), participants sat approximately 25 cm in front of a virtual table. The virtual table had the same height as the physical table, offering haptic feedback during the task. On the virtual table, a display presented a home position, a fixation cross, and a central circle surrounded by a series of other circles as the target. 

\subsubsection{Procedures and Design}

\begin{figure}
    \centering
    \includegraphics[width=1\linewidth]{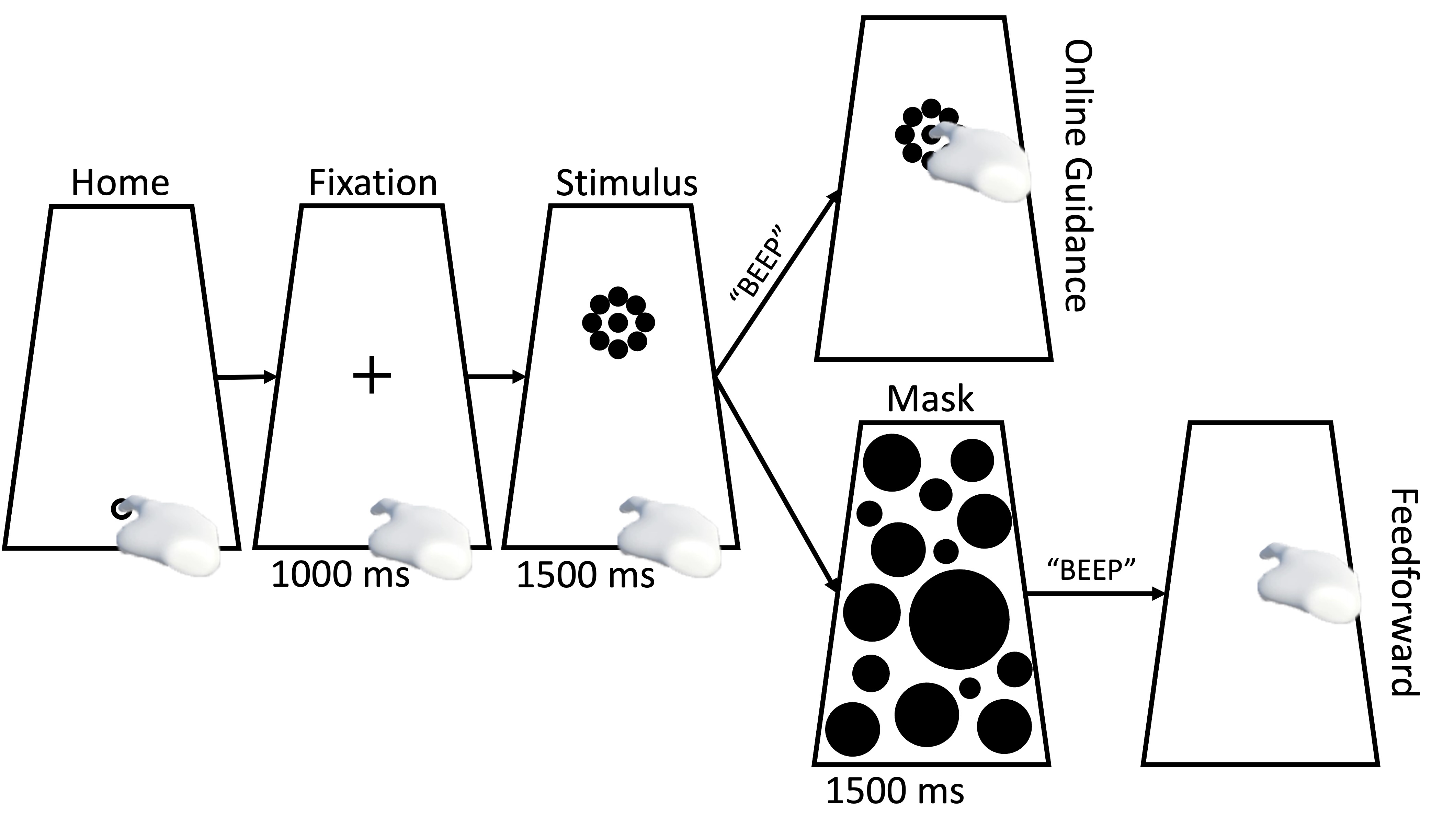}
    \caption{An illustration of the procedure for Experiment 1 from the participant’s point of view. The trapezoid represents the tabletop on which stimuli were displayed extending into the distance. The hand represents the virtual hand with which the participants performed the 3D pointing movement in VR. The participants could either see the target (Online Guidance) or not (Feedforward) when performing the movement after the beep.}
    \label{fig:exp1}
\end{figure}

For each trial, the participants were asked to perform a 3D pointing movement to the central circle of the target (Fig \ref{fig:exp1}). First, the participants moved their physical hand to the table so that the index finger of the virtual hand rested on the home position, 5 cm from the edge of the virtual table. When the virtual finger was on the home position, a fixation cross would appear for 1000 ms, followed by the target circles appearing for 1500 ms. In the online guidance condition, a beeping sound would play after the 1500 ms delay, and the participants would move to and land on the central circle as accurately and as quickly as they could. In the feedforward condition, the target circles would disappear and be replaced by a series of randomly sized and randomly placed circles as a mask. The mask would disappear after 1500 ms, followed by a beeping sound that prompted the participants to point to where they remembered the location of the central circle. In both conditions, the participants had a full vision of the virtual environment and the virtual hand throughout the entire experiment, including the sequence of each trial event. There were three potential distances between the home position and the target (20, 25, and 30 cm) and 48 repetitions for each feedback condition, resulting in a total of 288 trials. The experiment was blocked by feedback conditions, which were counterbalanced across participants.

\subsubsection{Data Analysis}

Kinematic data analysis was performed using TAT-HUM, a Python-based trajectory analysis toolkit for human movements \cite{wang2024tat}. 2.41\% of the total trials were removed due to false starts, slow movement, or missing data during data collection. For each trial, raw movement trajectories were filtered using a low-pass Butterworth filter with a sampling frequency of 250 Hz and a cutoff frequency of 10 Hz to remove high-frequency noise. Then, a central difference method was used to derive movement velocity. Movement onset and termination correspond to when the velocity along the depth axis exceeds or falls below a 50 mm/s threshold. The movement distance was the difference in displacement along the depth axis between movement termination and onset. Distance error (DE) for each movement was derived as the difference between movement and target distances, which was used to evaluate movement accuracy.

\subsubsection{Model Fitting}

DEs were fitted to the model based on Equation \ref{eq: const_err}, which contains two unknown parameters, $IPD$ and $\beta_\textrm{offset}$. The overall model fitting strategy was similar to that discussed in \cite{wang_geometry_2024}. Each participant's data from each feedback condition was divided based on a 70-30 train-test split. Two versions of the model were fitted. In the original version, one vergence offset $\beta$ and 20 participant-specific IPD values were fitted to the data (i.e., 21 parameters for the entire dataset). To anticipate the potential lack of effect of the vergence offset in one of the feedback conditions, an alternative model was also fitted where $\beta_\textrm{offset}$ was set to 0 (i.e., 20 participant-specific IPD values for the entire dataset). This alternative model was used to differentiate the impact of VAC on distance perception (feedforward) and disparity matching (online guidance). Model fitting was performed using SciPy \cite{virtanen_scipy_2020}, optimized with non-linear least squares with the Levenberg-Marquardt algorithm \cite{more2006levenberg}. Bayesian information criterion (BIC) and $r^2$ were used to evaluate the model's goodness of fit. 

\subsection{Results}

\begin{table}[h]
\centering
\begin{tabular}{clllll}
\hline
\multicolumn{1}{l}{}             &           & \multicolumn{2}{c}{BIC} & \multicolumn{2}{c}{$r^2$} \\ \cline{3-6} 
\multicolumn{1}{c}{Feedback}     & Model     & Train      & Test       & Train      & Test       \\ \hline
\multirow{2}{*}{Online Guidance} & \bf{Original}  & \bf{21.13}  & \bf{73.89}  & \bf{0.80} & \bf{0.61} \\
                                 & No Offset & 76.96  & 113.03 & 0.46 & 0.20 \\ \hline
\multirow{2}{*}{Feedforward}     & Original  & 112.54 & 148.18 & 0.94 & 0.90 \\
                                 & \bf{No Offset} & \bf{108.87} & \bf{144.23} & \bf{0.94} & \bf{0.90} \\ \hline \\
\end{tabular}
\caption{Comparison of BIC and $r^2$ values for different model versions and feedback conditions.}
\label{tab:model_comparison}
\end{table}

 Table \ref{tab:model_comparison} shows the goodness-of-fit measures for different model versions and different feedback conditions. Similar to \cite{wang_geometry_2024}, the models provided a good fit for the data. For the online guidance condition, the original version yielded noticeably smaller BIC values for both the train and test data and the fitted model could account for 61\% of the variance in the test data. Importantly, the fitted vergence offset equals 0.23°, consistent with the value of 0.22° that was reported in \cite{wang_geometry_2024}. For the feedforward condition, while the two versions of the model yielded similarly high $r^2$ values, the no vergence offset version had smaller BIC values due to having one fewer parameter. Overall, the model comparison suggests that the vergence offset is appropriate to describe DE if the movement was performed with online visual guidance when the target was present during the movement, but not in a feedforward manner when the actor needs to rely on the visual memory of the target's location.

\begin{figure}
    \centering
    \includegraphics[width=0.9\linewidth]{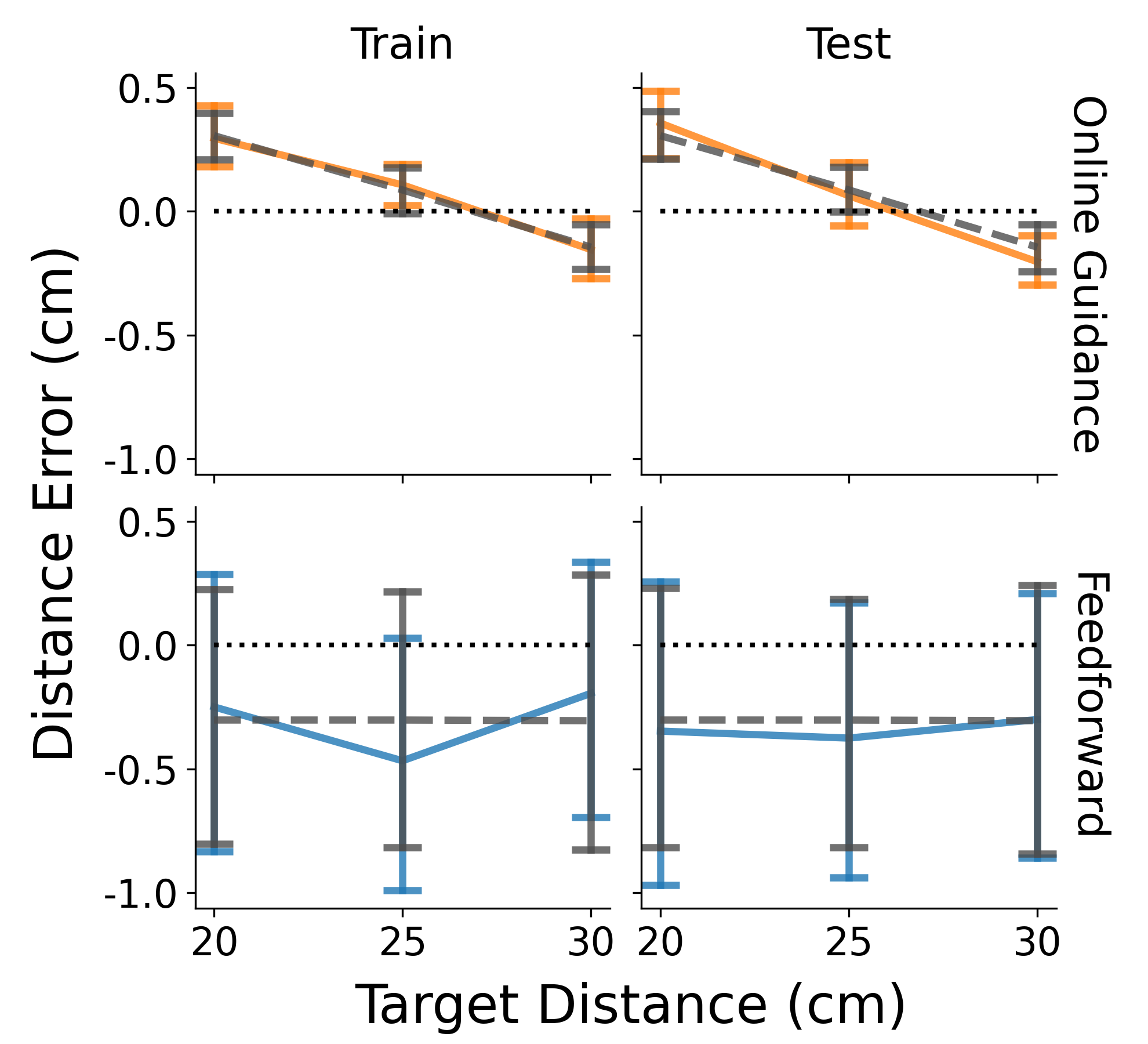}
    \caption{The mean distance errors from the behavioral experiment (solid lines) and model predictions (gray, dashed lines) as a function of target distance for different feedback conditions in Experiment 1. Model predictions were based on the optimal versions of the model for each feedback condition. Error bars represent 95\% confidence intervals. }
    \label{fig:ebbinghaus_fit}
\end{figure}

Fig. \ref{fig:ebbinghaus_fit} shows the mean DE from the behavioral experiment and model fitting for different feedback conditions. Consistent with the goodness-of-fit measures, the model fitted the behavioral results relatively well. A repeated-measures analysis of variance (ANOVA) was performed to evaluate the effects of target distance and feedback on DE. There was a significant effect of target distance, $F(1.38, 26.17) = 9.29, p < 0.01, \eta^2_p = 0.33$, but not feedback $F(1, 19) = 2.16, p = 0.16, \eta^2_p = 0.10$. The interaction was significant, $F(1.45, 27.58) = 12.96, p < 0.001, \eta^2_p = 0.41$, which is of particular interest. The online guidance condition demonstrated a downward trend in DE as a function of target distance. This suggests an increasing depth compression as distance increases. Post-hoc pairwise comparisons showed a significant difference between all distances (20 - 25: $t(19) = 4.76, p < 0.001$, 25 - 30: $t(19) = 5.67, p < 0.001$; 20 - 30: $t(19) = 8.08, p < 0.001$). These findings are consistent with those reported in \cite{wang_geometry_2024}. With a different experimental setup and different participants, the successful replication confirmed the validity of the geometrical model for VAC, suggesting that VAC draws the users’ fixation inward, thus perturbing the binocular viewing geometry and affecting the accuracy of targeted movement. 

Interestingly, DE in the feedforward condition remained relatively stable across distances, albeit more variable. Post-hoc pairwise comparisons revealed no significant differences between most target distances (20 - 25 cm: $t(19) = 1.69, p = 0.24$; 20 - 30 cm: $t(19) = 0.40, p = 0.92$), except for between 25 and 30 cm ($t(19) = 3.11, p < 0.05$). Originally, the downward trend in DE as a function of target distance is indicative of the effect of VAC on targeted movement. With feedforward movement, the modulating effect of target distance is no longer present. Combining this insight with the model-fitting result, it is concluded that VAC and the vergence offset primarily affect the disparity-matching process during online control of movement when both hand and target are visible throughout the movement.

Why does the vergence offset affect disparity matching for online movement, but not memory-based feedforward movement? In goal-directed movements, humans integrate various sources of sensory information (e.g., visual, proprioceptive, kinesthetic, cutaneous) to plan and execute the movements \cite{sober2003multisensory, sober2005flexible}. In this process, the hand's position is estimated through vision and proprioception, which are suggested to be integrated optimally \cite{ghahramani1997computational} based on their respective reliability (or the inverse of their variances) \cite{bays2007computational, ernst2002humans, liu2018spatial}. During online movement, the visual system continuously processes binocular disparity for both the hand and target, enabling real-time corrections based on their relative position \cite{anderson_solution_2010}. This process is stable \cite{bingham2023stable}, leading to a greater weighting of visual information during online movement control. As the movement progresses, the hand eventually occludes the target, making further disparity matching impossible. At this point, the actor must rely on feedforward mechanisms, using the last available relative disparity information before occlusion and proprioceptive information to guide the final portion of the movement. Since the vergence offset perturbs visual information but not proprioceptive information, a mismatch occurs between the online (visual) and feedforward (proprioceptive) phases of the movement, leading to movement endpoint errors due to conflicting spatial information. 

In summary, Experiment 1 replicated the results in \cite{wang_geometry_2024} and, importantly, demonstrated that the vergence offset primarily affects movements guided using online visual information. After further validating the effectiveness of the geometrical VAC model, attention could be directed to investigating whether such a mechanistic understanding of the process could be applied to ameliorate the effect of VAC on targeted movement in VR.

\section{VAC Transformation Algorithm} \label{vac_algo}

While completely eliminating the impact of VAC on the user's perceptuomotor experience could be challenging, the geometrical description of VAC reveals a potential solution that optimizes the user’s perceptuomotor experience around VAC. On the rendering side, the VR HMD’s displays provide consistent disparity information. On the perceptual side, VAC affects how the rendered disparity is interpreted, where the vergence offset leads to a distorted viewing geometry. Since the vergence offset can be empirically derived, a pragmatic solution is to incorporate this offset within the rendering process. Specifically, the vergence offset can be transferred to the target’s visual angle:
\begin{equation}\label{eq: offset_swap}
    \delta = (\phi + \beta_\textrm{offset}) - \tau = \phi - (\tau - \beta_\textrm{offset})
\end{equation}

This approach entails systematically rendering the objects at a farther distance relative to the user’s cyclopean eye (i.e., the VR camera), in which case these objects’ perceived locations would be consistent with those specified by the perturbed viewing geometry. That is, if the vergence offset elicits movement undershooting, the target’s location can be presented as being slightly further so that when moving towards the target, the users would land on the location of the target.

Based on Equation \ref{eq: offset_swap}, the transformed target should yield a visual angle of $\tilde{\tau}$ (Fig. \ref{fig:vac_demo}c):
\begin{equation*}
    \tilde{\tau} = \tau - \beta_\textrm{offset}   
\end{equation*}
The distance between the transformed target point and the cyclopean eye is:
\begin{equation*}
    \tilde{d} = \dfrac{IPD / 2}{\tan{(\tilde{\tau}} / 2)}
\end{equation*}
Because VAC only affects depth perception, but not dimensions along the $x$ and $y$ directions, the transformed $z$-coordinate should be:
\begin{equation}
    \tilde{z} = \sqrt{\tilde{d} ^ 2 - x^2_p - y^2_p}
    \label{eq: transformedZ}
\end{equation}
where $x^2_P$ and $y^2_P$ are the $x$- and $y$-coordinates of the target. 

In practice, the relationship in \eqref{eq: transformedZ} can be implemented as a shader program that systematically transforms the rendered virtual environment. Specifically, in Unity, this transformation can be implemented as a replacement shader attached to the VR camera. Because the binocular viewing geometry is measured based on the positions of the eyes, the transformation has to be performed in the view space in the vertex shader, where the viewing distance can be extracted as the $z$-coordinate, $z_\textrm{view}$. With the HMD's $IPD_\textrm{HMD}$, the visual angle of a point can be computed:
\begin{equation}\label{eq: visual_angle}
    \theta = \textrm{atan2}(IPD_\textrm{HMD} / 2, z_\textrm{view})
\end{equation}
The updated visual angle for this point that factors in the vergence offset is
\begin{equation}\label{eq: visual_angle_offset}
    \tilde{\theta} = \theta - \beta_\textrm{offset}
\end{equation}
Therefore, the transformed $\tilde{z}_\textrm{view}$ is 
\begin{equation}\label{eq: shader_z}
    \tilde{z}_\textrm{view} = \dfrac{IPD_\textrm{HMD} / 2}{\tan(\tilde{\theta})}
\end{equation}
which can be transformed back into the world space and assigned to the original vertex. 

\begin{figure}
    \centering
    \includegraphics[width=0.7\linewidth]{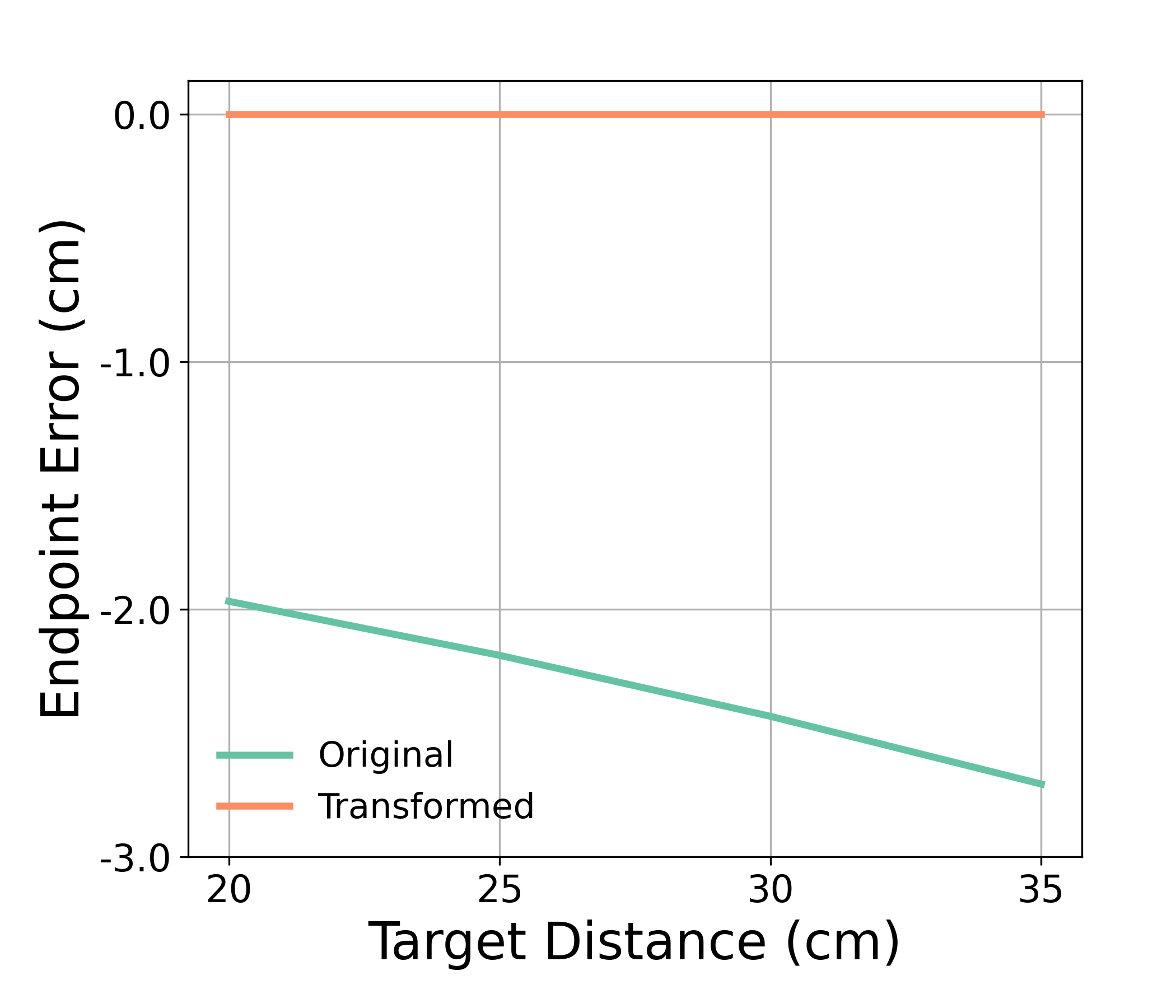}
    \caption{Predicted endpoint errors for the original and transformed virtual environment based on Equation \ref{eq: shader_z} as a function of target lengths.}
    \label{fig:validation prediction}
\end{figure}

Fig. \ref{fig:validation prediction} shows the predicted endpoint errors for the original and transformed virtual environments. Note that because Experiment 1 showed that VAC affects disparity matching in online movements, endpoint errors, instead of DE, were visualized to capture the accuracy of disparity matching. For the original environment, the endpoint errors indicate consistent undershooting, which increases with a longer target distance. This is different from DE (e.g., Fig. \ref{fig:ebbinghaus_fit}), which demonstrates distance overestimation and underestimation depending on target distance. For the transformed environment, the endpoint errors are at 0 and remain consistent across target distances, indicating the transformation enables accurate movement endpoint.

\section{Experiment 2: Validation}

\begin{figure}
    \centering
    \includegraphics[width=1\linewidth]{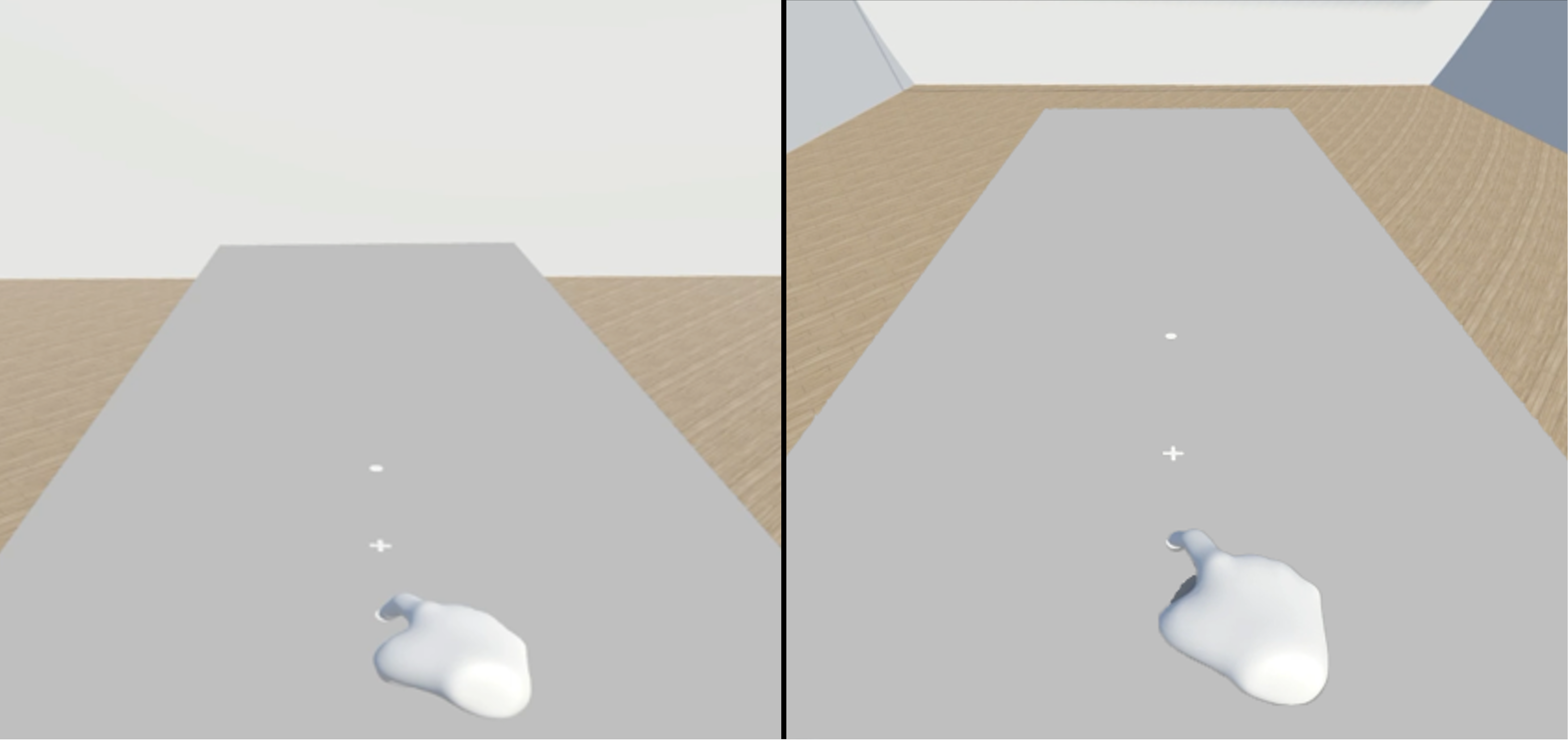}
    \caption{The participant's view through the HMD during Experiment 2 without (left) and with (right) the transformation applied to the virtual environment. }
    \label{fig:vr_transformation_demo}
\end{figure}

Experiment 2 was designed with a similar setup as Experiment 1 to test the effectiveness of the transformation algorithm on ameliorating movement undershooting in VR. To test this, participants performed a series of 3D pointing movements with or without the transformation shader enabled (Fig. \ref{fig:vr_transformation_demo}). The shader used the empirically-derived vergence offset, $\beta_\textrm{offset} = 0.22^{\circ}$ (Experiment 1 and \cite{wang_geometry_2024}). Because Experiment 1 showed that VAC only affects online but not feedforward movement, visual feedback was always available.

\subsection{Methods}

\subsubsection{Participants}

Twenty-three (23) adults participated in this experiment (12 females and 11 males, between 19 and 32 years of age). All participants were right-handed, had normal or corrected-to-normal vision, and were free from any known neurological impairments. Participants provided full and informed consent before taking part in the study either as volunteers or receiving monetary compensation.

\subsection{Materials and Apparatus}

Experiment 2 was designed using the same equipment (i.e., HTC VIVE Pro Eye and Optotrak) and virtual environment as in Experiment 1. The only difference with Experiment 1 was that the participants were only presented with a single circle as the target. The transformation algorithm was implemented as a replacement shader in Unity as described in Equations \ref{eq: visual_angle}-\ref{eq: shader_z}. 

\subsubsection{Procedures and Design}

The procedures of this experiment were similar to those of Experiment 1. The target circle could appear 20, 25, 30, or 35 cm away from the home position. There were two visual conditions (original and transformed) and only one feedback condition (online guidance). Because the human visual system could adapt to VAC by decoupling the vergence and accommodation processes \cite{eadie_modelling_2000, wang_prolonged_2024}, this experiment was blocked by visual conditions, randomized between participants. There were 16 repetitions, resulting in 2 (visual condition: original, transformed) $\times$ 4 (target distances: 20, 25, 30, 35 cm) $\times$ 16 repetitions = 128 trials.

\subsubsection{Data Analysis}

Raw movement trajectory data were processed as in Experiment 1 using TAT-HUM \cite{wang2024tat}. The disparity difference between the hand and the target was calculated to evaluate the accuracy of disparity matching. Given the target location, the target and hand’s visual angles, $\tau$ and $\eta$, can be derived using Equation \ref{eq: visual_angle}. Because the fixation angle cancels out (see Equation \ref{eq: delta}), the disparity difference between the hand and the target is $\eta - \tau$. Endpoint errors were also derived to interpret the disparity difference, which was the difference between the target and endpoint locations along the depth axis.

\subsection{Results}

\begin{figure*}[!htbp]
    \centering
    \includegraphics[width=0.8\linewidth]{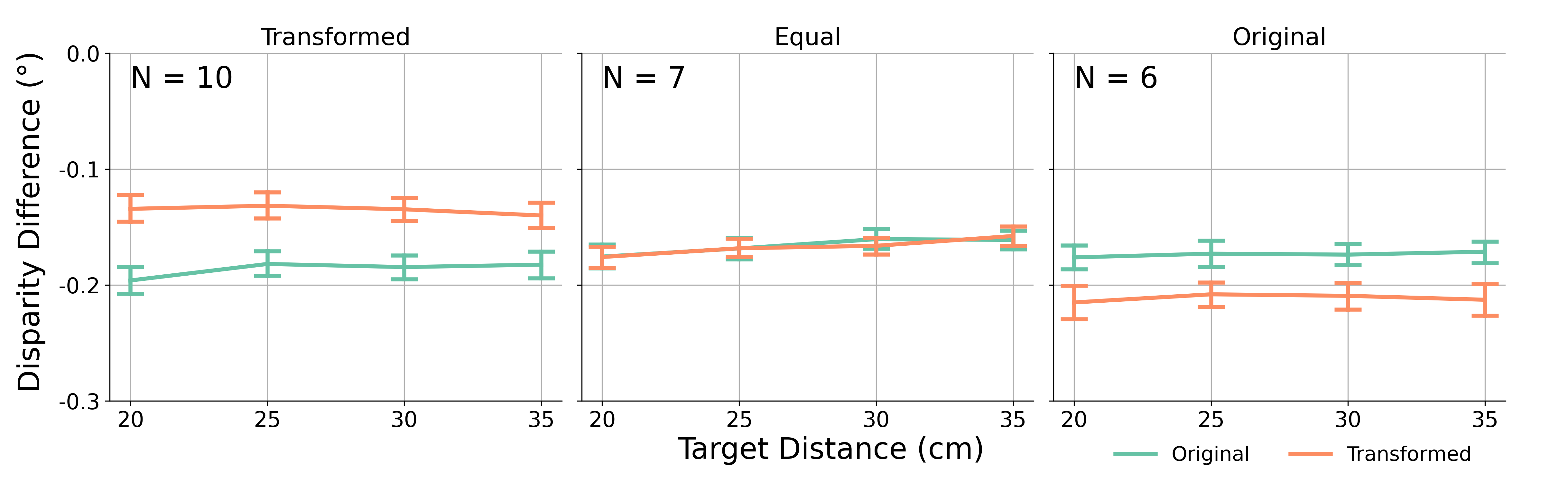}
    \caption{The mean disparity difference between the hand and the target as a function of target distance for different visual conditions and for different groups of participants. Error bars represent 95\% confidence intervals.}
    \label{fig:validation_disparity_diff}
\end{figure*}

Preliminary analysis showed noticeable individual differences in terms of how individuals responded to the different visual conditions. To account for the individual differences, a series of $t$-tests was used to compare the disparity difference between the original and transformed conditions for each participant. Based on the t-test results, participants were separated into three groups: Transformed, Equal, and Original. Individuals in the Transformed group had more accurate performance (i.e., disparity difference closer to 0) in the transformed condition than in the original condition (N = 10, or 44\%). Those in the Original group had more accurate performance in the original condition (N = 7, or 30\%). Finally, participants in the Equal group did not yield a significant difference in performance between the original and transformed conditions (N = 6, or 26\%). Fig. \ref{fig:validation_disparity_diff} shows the mean disparity difference for different participant groups. Across the three groups, the original condition yielded consistent disparity difference across different target distances, at around 0.17°. In contrast, different participant groups responded to the transformed condition differently.

\begin{figure*}[!htbp]
    \centering
    \includegraphics[width=0.8\linewidth]{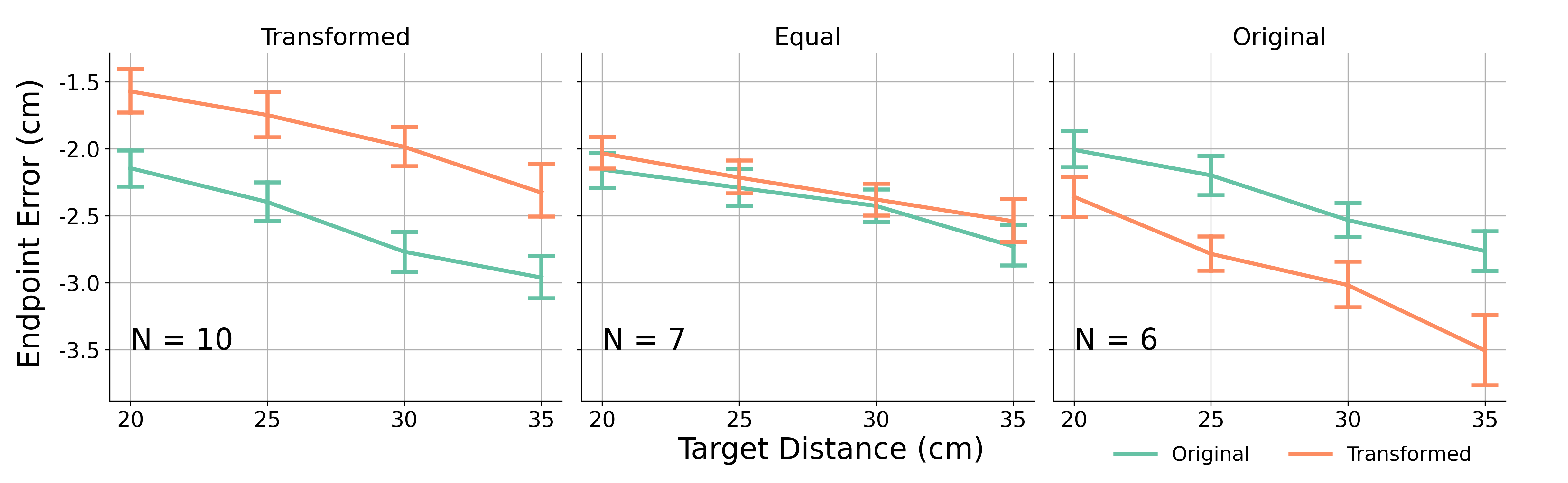}
    \caption{The mean endpoint errors as a function of target distance for different visual conditions and for different groups of participants. See the text for how each group was determined. }
    \label{fig:validation_ce}
\end{figure*}


A mixed-design ANOVA on disparity difference with two within-subject factors (visual condition, target distance) and one between-subject factor (participant group: Transformed, Equal, and Original) was conducted. The only significant effect involving visual condition was the interaction between the participant group and visual condition, $F(2, 20) = 17.40, p < 0.001, \eta^2_p = 0.64$. The main effect of visual condition was not significant, $F(1, 20) = 0.40, p = 0.54, \eta^2_p = 0.020$, nor was the interaction between visual condition and target distance, $F(3, 60) = 0.17, p = 0.92, \eta^2_p = 0.008$.

Post-hoc pairwise comparisons showed that none of the participant groups had different disparity differences in the original condition (Transform-Equal: $t(20) = 1.67, p = 0.24$; Transform-Original: $t(20) = 1.00, p = 0.58$; Original-Equal: $t(20) = 0.54, p = 0.85$). The average disparity difference for the original condition across all participant groups was -0.19°, approximately the values for the derived vergence offset (-0.22°). Moreover, each participant group yielded different disparity differences for the original and transformed conditions. For the Transformed group, the disparity difference was closer to 0 in the transformed condition than in the original condition (mean difference = 0.05°). Because the goal of disparity matching is to drive the disparity corresponding to the hand and the target to 0, the Transformed group demonstrated a 29\% improvement in the accuracy of disparity matching, suggesting the effectiveness of the algorithm in facilitating the disparity matching process given VAC. In contrast, for the Original group, the magnitude of the disparity difference increased by 0.03° in the transformed condition, or a 16\% performance decline. Finally, the Equal group did not yield any difference between the two conditions.

Curiously, target distance did not have any significant effect, including the main effect ($F(2.00, 39.92) = 1.70, p = 0.20, \eta^2_p = 0.078$) and interaction effects (participant group $\times$ target distance: $F(3.99, 39.92) = 0.57, p = 0.68, \eta^2_p = 0.054$; visual condition $\times$ target distance: $F(3, 60) = 0.17, p = 0.92, \eta^2_p = 0.008$; participant group $\times$ visual condition $\times$ target distance: $F(6, 60) = 0.29, p = 0.94, \eta^2_p = 0.028$). As demonstrated in \ref{fig:validation_disparity_diff}, the disparity difference remained relatively stable across all conditions and participant groups. This finding is in stark contrast to the negative trend between target distance and distance-based errors observed in previous experiments.

Fig. \ref{fig:validation_ce} shows the equivalent endpoint errors for different participant groups. Noticeably, different from the previous analysis with the distance-based measure (Fig. \ref{fig:ebbinghaus_fit}), the movement endpoint demonstrated consistent undershooting (mean $\approx$ -2.22 cm). Moreover, unlike the disparity difference, the negative trend between movement error and target distance was present here. The same ANOVA on endpoint error showed a significant main effect of target distance, $F(2.12, 42.31) = 65.21, p < 0.001, \eta^2_p = 0.77$. As shown in Fig. \ref{fig:validation_ce}, the endpoint errors increased as a function of target distance, indicating an increased movement undershooting. Similar to the disparity difference, there was a significant interaction between the participant group and visual condition, $F(2, 20) = 12.28, p < 0.001, \eta^2_p = 0.55$. For the Transformed group, the transformed condition reduced the endpoint error by 0.73 cm (or a 28\% improvement). In comparison, the Original group elicited a 0.42 cm (or a 14\%) increase in endpoint error. Finally, the ANOVA did not show any other significant effects, including the interaction between visual condition and target distance ($F(3, 60) = 0.15, p = 0.93, \eta^2_p = 0.007$). Endpoint errors did not exhibit the predicted outcome shown in Fig. \ref{fig:validation prediction}. Instead, there was a uniform improvement in endpoint accuracy across all target distances. 

\section{Discussion}

The current work sought to address one of the critical challenges in VR technology - the vergence-accommodation conflict (VAC). Through the development and validation of a human vision-based geometrical model, this work provides new insights into how VAC affects targeted movements in VR and offers a software-based solution to mitigate the impact of VAC. The results from Experiment 1 confirmed the predictions of the geometrical model, demonstrating that VAC induces a constant offset in the vergence angle that leads to systematic movement undershooting during 3D pointing tasks. This effect was most pronounced during online guidance, where real-time visual feedback of both the target and the effector is essential for accurate movement execution. The finding that VAC primarily affects online, rather than feedforward movements, suggests that the continuous processing of binocular disparity is particularly vulnerable to the disruptions caused by VAC. This finding highlights the importance of addressing VAC in VR systems that rely heavily on real-time interaction. 

Moreover, by successfully implementing a shader program based on the geometrical model, the results of Experiment 2 showed that this software solution could improve movement accuracy by approximately 30\%. This advancement not only demonstrates the practical viability of the model but also offers a potential tool for VR developers to enhance user experience in HMD-based applications. The observed individual differences in the effectiveness of the algorithm suggest that the impact of VAC may vary depending on factors such as user-specific visual characteristics or task-specific demands. These findings could inform the customization of VR experiences to better accommodate individual users, potentially leading to more personalized and effective VR systems. 

One aspect of the findings is worthy of extra attention - the algorithm's ineffectiveness of reducing the increasing undershooting at farther target distances. Combining this observation with results from Experiments 1 and 2, it can be suggested that there is a potential interaction between vergence offset and target distance that plays a pivotal role in the increasing undershooting. Evidence supporting this conclusion includes the observation that the VAC's impacts were only evident in the online guidance condition, not in the feedforward condition. As noted, online movement control includes a feedforward component near the end of the movement when the hand occludes the target and when visuomotor delays preclude further online correction. However, online error detection and correction earlier in the movement rely on the perceived disparity between the hand and the target. This feedforward process at the end of the movement allows for accurate targeting despite occlusion in a natural viewing condition. However, in VR, it increases undershooting, suggesting that the disparity information before occlusion inaccurately represents the spatial relationship between the hand and the target. This discrepancy indicates that the vergence offset affects the perceived locations of the hand and target differently, depending on their distances. To explore this hypothesis further, future studies could investigate how hiding the hand at various points during the movement impacts endpoint accuracy. 

Furthermore, the contrasting patterns between endpoint- and angle-based measures in Experiment 2 also indicate the interaction between vergence offset and target distance. While the endpoint indicates increasing undershooting as the target gets farther from the participant, the disparity difference between the target and the hand remains constant across different target distances. In the original condition, the disparity difference was similar to the vergence offset ($\approx$ 0.22°), at around 0.19°. However, this constant angular offset translates to a downward-trending endpoint error because of the nonlinear mapping between visual angle and distance. Therefore, flattening the systematic variations of endpoint error as a function of target distance may require setting different vergence offset values across different target distances. The next iteration of the model could introduce different vergence offsets to targets at different distances, which may help to make the endpoint errors consistent across different target distances. 

Compared to other hardware and software solutions to VAC, the solution presented in the current study leverages the intrinsic binocular viewing geometry of VR HMDs to present a more accessible and potentially cost-effective alternative. While previous studies have explored various strategies to address VAC, including hardware-based solutions such as adaptive optics \cite{laffont2018verifocal} and software-based solutions such as simulated blur \cite{hussain2023improving}, these solutions have yet to see widespread adoption due to their complexity and cost. The shader-based approach presented herein, which can be implemented within existing VR systems without additional hardware, presents a practical and scalable solution to this persistent problem. The improvement in movement accuracy aligns with findings from other studies that emphasize the importance of addressing VAC for enhancing user performance in VR.

\subsection{Limitations}

Despite these promising results, the current study contains several limitations that warrant additional studies. First, while the algorithm aims to improve movement accuracy despite VAC, the algorithm itself does not factor in the effect of a lack of blurring in the process. The HMD’s displays present everything in sharp focus, which is different from the natural viewing experience where only foveal vision is focused. This difference may further contribute to the decoupling between vergence and accommodation, disrupting the binocular viewing geometry. Therefore, it could be worthwhile to incorporate other algorithms, such as \cite{hussain2023improving}, in order to further improve task performance.

Second, while this study identified noticeable individual differences in how the participants perform in the transformed virtual environment, it did not offer any relevant evidence to delineate the source of such individual differences. These variations could be attributed to individual differences in visual acuity, cognitive processing, or prior experience with VR technology. Future research should aim to explore these potential sources of variability by incorporating a more detailed analysis of participant characteristics, including visual tests, cognitive assessments, and VR usage history. Understanding these factors could provide deeper insights into the effectiveness of transformation algorithms and help tailor VR experiences to better accommodate diverse user needs.


\section{Conclusion}

The work reported herein was designed to investigate the impact of VAC on users’ perceptual geometry and the accuracy of 3D pointing movements in VR. The first study confirmed that VAC draws the vergence angle inward, shortening the fixation distance and distorting binocular viewing geometry, leading to movement undershooting. Notably, the study demonstrated that VAC primarily affects movements performed with online visual feedback, rather than memory-based feedforward movements. This finding suggests that VAC disrupts the disparity matching process during online movement control, but does not significantly impair distance perception for planning feedforward movements.

Building on this geometrical understanding of VAC, a transformation algorithm was developed to mitigate its effects on movement accuracy. The second study showed that the algorithm improved movement accuracy by approximately 30\%. However, the algorithm’s effectiveness varied among participants; for some, it had little to no effect or even slightly degraded performance. Furthermore, while the algorithm consistently enhanced disparity matching accuracy, this improvement did not alleviate the increasing undershooting as target distance increased. Therefore, further studies are needed to investigate these discrepancies. Finally, insight from the current study has significant implications for improving the design of VR systems in applications where real-time precision is critical, such as surgical training and remote robotic control. By better understanding how VAC disrupts online movement control, developers can refine VR systems to enhance accuracy in tasks requiring fine motor control and precise depth perception.

\bibliographystyle{ieeetr}
\bibliography{references}

\end{document}